\documentclass[reprint,twocolumn,amsmath,amssymb,aps,prb]{revtex4}

\usepackage{hyperref}
\usepackage[english]{babel} 
\usepackage{amssymb}
\usepackage{amsmath}
\usepackage{txfonts}
\usepackage{mathdots}
\usepackage[classicReIm]{kpfonts}
\usepackage{graphicx}
\usepackage[dvipsnames]{xcolor}
\def\arrvline{\hfil\kern\arraycolsep\vline\kern-\arraycolsep\hfilneg}
\usepackage{dcolumn}
\usepackage{bm}

\begin{document}

\preprint{APS/123-QED}

\title{Electrical control of excitons in GaN/(Al,Ga)N quantum wells}
\author{R.~Aristegui, F.~Chiaruttini, B.~Jouault, P.~Lefebvre, C.~Brimont, T. Guillet, M. Vladimirova}
\affiliation{Laboratoire Charles Coulomb (L2C), University of Montpellier, CNRS, Montpellier, France}

\author{S. Chenot, Y. Cordier, B. Damilano}
\affiliation{{CRHEA, Universit\'e C\^ote d’Azur, CNRS, Valbonne, France}
}%

\begin{abstract}
  {A} giant built-in electric field in the growth direction makes excitons in wide GaN/(Al, Ga)N quantum wells spatially indirect even in the absence of any external bias. 
%
Significant densities of indirect excitons  can accumulate in electrostatic traps  imprinted in the quantum well plane by a thin metal layer deposited on top of the heterostructure.
%
By jointly measuring  spatially-resolved photoluminescence  and   photo-induced current, we demonstrate that exciton density in the trap can be controlled via an external electric bias, which is capable of altering the trap depth. 
Application of a negative bias  deepens the  trapping potential, but does not lead to any additional  accumulation of excitons in the trap. This is  due to exciton dissociation instigated by the lateral  electric field at the electrode edges. The resulting carrier losses are detected as an increased photo-current and  reduced photoluminescence intensity.
By contrast, application of a positive bias washes out the electrode-induced trapping  potential. 
Thus, excitons get released from the trap and recover free propagation in the plane that we reveal by spatially-resolved photoluminescence.
\end{abstract}

\maketitle

\section{Introduction}
\label{sec:Introduction}

Indirect excitons (IXs) are  bound pairs of an electron and a hole confined in separated quantum layers, either semiconductor quantum wells (QWs) \cite{Chen1987,Alexandrou1990}, or various transition metal dichalcogenides (TMD) monolayers \cite{Rivera2015,Gerber2019,Unuchek2018,Jauregui2019}. Due to their  permanent dipole moment, IXs can be controlled in-situ by voltage \cite{High2007,Mak2018,Jauregui2019}, can travel over large distances \cite{Voros2005,Rapaport2006,Dorow2018,Unuchek2018,Wagner2021,Sun2022}, and can cool below the temperature of quantum degeneracy  before recombination \cite{High2012,Anankine2017,Cohen2016,MazuzHarpaz2017,Shilo2013,Stern2014,Misra2018,Schinner2013,Butov2016,Combescot2017,Wang2019}. Due to these properties, IXs are considered as a promising platform for the development of excitonic devices \cite{Butov2017,Liu2022}. Most prominent,  experimentally-demonstrated devices operating via electrical bias control are based on GaAs. These  include traps, lattices, and excitonic transistors \cite{Huber1998,Hammack2006,High2007,High2008,Andreakou2014,Remeika2015}.

Excitons in {  
{GaN/(Al,Ga)N QWs grown along the ($0001$) crystal axis}
} 
are naturally indirect, as their electron and hole  constituents  are spatially separated in growth direction due to a strong built-in electric field \cite{Leroux1998,Grandjean1999,Gil2014,Fedichkin2015,Fedichkin2016}. As a result, GaN-based QWs differ from other heterostructures  hosting indirect excitons -- such as GaAs-based double QWs and recently developed TMD heterobilayers --  where an external electric bias is usually applied to push the IX transition energy below that of the direct exciton \cite{Chen1987,Zrenner1992,Calman2016,Jauregui2019}.

\begin{figure}[t!]
	\includegraphics[width=3.4in]{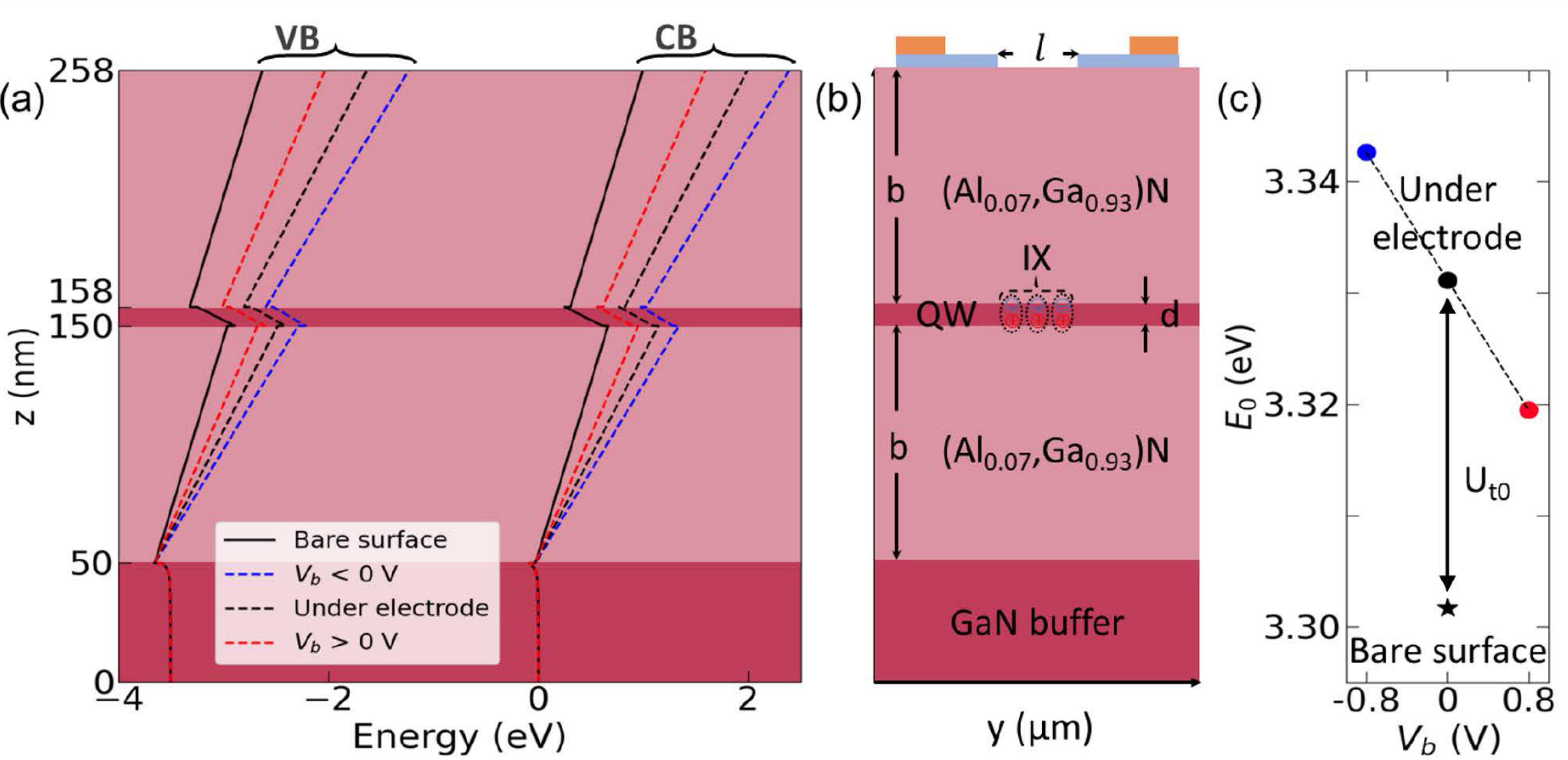}
	\caption{  (a) Calculated  valence (VB) and conduction (CB) band diagrams for the studied sample in the absence (solid line) and in the presence (dashed lines) of the electrodes on the surface. (b) yz-view of the sample covered by semi-transparent  ($5$~nm Au/$5$~nm Ni, blue)  and opaque ($200$~nm Au, orange) electrodes. (c) Calculated zero-density exciton energy in the electrode-covered regions at different voltages (circles) and in the  bare surface areas (star). Arrow shows $U_{t0}$, the zero-density trapping potential depth,  at $V_b=0$~V.
	}
		\label{fig:fig1}
\end{figure}

We have shown previously that  in-plane confinement and cooling of GaN-hosted IXs can be achieved  without any applied electric bias  via  an electrostatic potential created by metallic patterns, of various shapes, deposited on the sample surface\cite{Chiaruttini2019}. In contrast with GaAs and TMD-based systems, here IXs accumulate in {
{ the bare surface regions, rather than in the metal-covered areas (see Fig.~\ref{fig:fig1}~(b)). This amends the  collection of light emitted by trapped IXs, and could be potentially advantageous for IX coherence, since  spurious electrons that inevitably emerge in the presence of an electric bias are known to be detrimental for trapped IX coherence properties \cite{Honold1989,Koch1995,Anankine2018}}
}.

In this work we start to exploit these advantages. IXs that we study are  created optically in the plane of GaN QW. The sample surface is patterned with electrodes similar to those realized in Ref.~\onlinecite{Chiaruttini2019}, but here we apply to them an external  electric bias, and detect both spatially-resolved emission and the corresponding photo-current.
This allows us to demonstrate that (i) IXs can be released from the trap upon application of a positive bias due to reduction of the trap depth; (ii) application of a negative bias does increase the  depth of the trap but fails augmenting IX accumulation;  (iii)  the efficiency of the IX density control by the external bias is limited by the increasingly high in-plane component of the electric field, that leads to IX dissociation in the presence of a strong, negative bias.

\begin{figure}
	\includegraphics[width=3.4in]{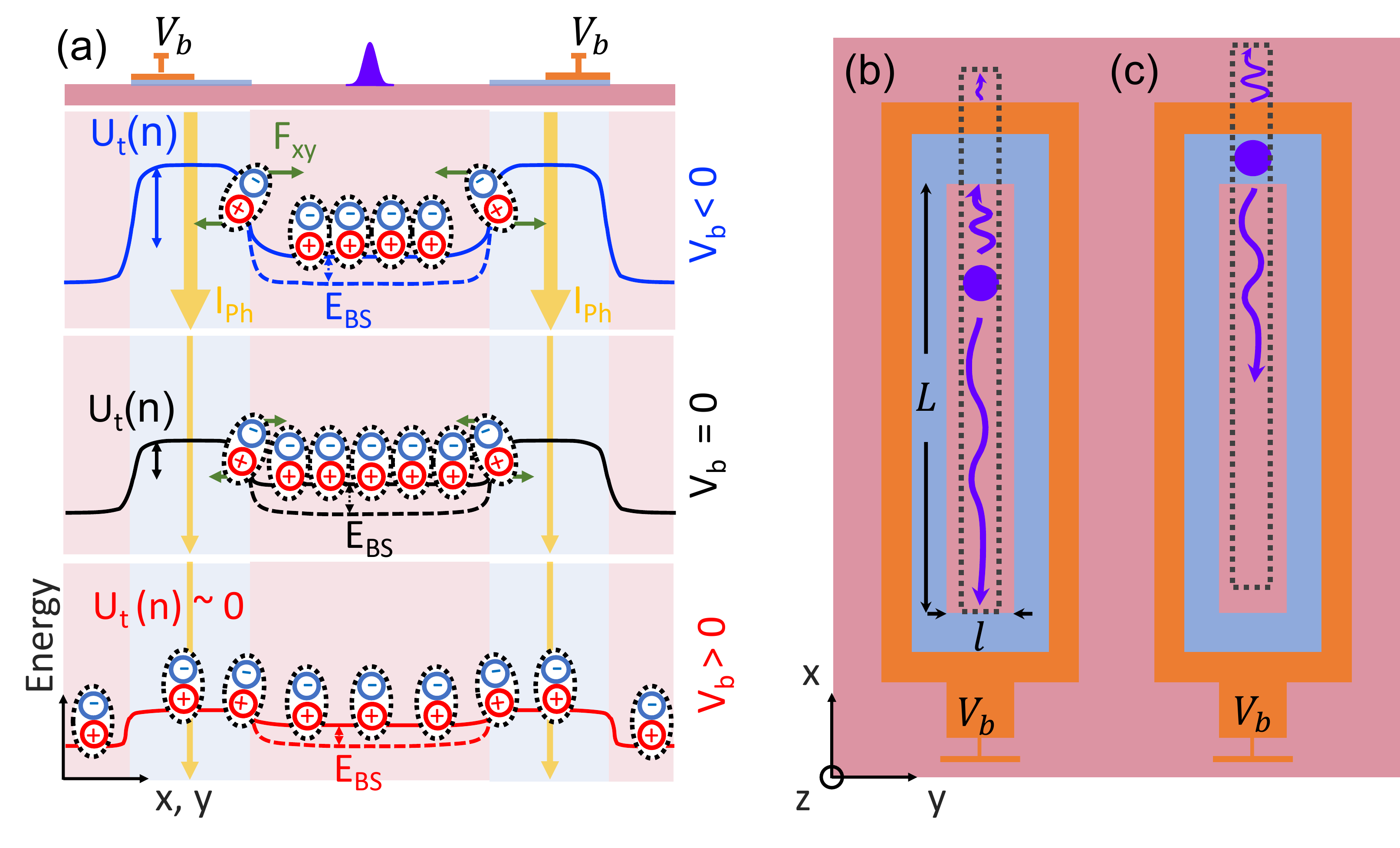}
	\caption{(a) Sketch of the IX energy profile along x(y)-axis at different applied biases $V_b$ in the zero-density regime (dashed lines) and under optical pumping (solid lines).  IX population, trap depth $U_t(n)=U_{t0}-E_{BS}$, IX blueshift $E_{BS}$, lateral component of the electric field $F_{xy}$  at the trap edges and photo-current $I_{ph}$ are schematically represented. (b-c)  xy view of the sample surface covered by electrodes and two experimental geometries: excitation on the bare surface within the trap (b) and through the semi-transparent electrode (c). Semi-transparent (opaque)  electrodes are shown in blue (orange).  Dashed frames indicate the corresponding detection areas.
	}
		\label{fig:fig2}
\end{figure}

\section{Samples and experimental setup}
\label{sec:samples}
%

%
{
{The studied sample is 
 grown by molecular beam epitaxy on a free-standing (0001)-oriented GaN substrate (AMMONO, threading dislocation density $\approx 10^4$~cm$^{-2}$), followed by 
$1200$~nm-thick GaN buffer. It contains a wide  GaN QW ($d=8$~nm) sandwiched between two identical
 Al$_{0.07}$Ga$_{0.93}$N barriers ($b=100$~nm), see Fig.~\ref{fig:fig1}~(b). }}

The built-in electric field in the growth direction that we estimate as $F_z \approx 450$~kV/cm \cite{nextnano} makes excitons in this QW spatially indirect. The electrons are pushed by this field towards the sample surface, while the holes move towards the substrate; this is the quantum-confined Stark effect \cite{Leroux1998}. As a consequence, in the zero-density limit, IX energy is as low as $E_0 \approx 3.3$~eV, that is about $0.18$~eV below GaN bandgap at $10$~K. The corresponding radiative lifetime is in the order of microseconds, binding energy  $E_b=18.6$~meV, and dipole length is close to the QW width \cite{Fedichkin2016,Chiaruttini2019,Chiaruttini2021}.

In order to create an in-plane trapping potential, semi-transparent electrodes consisting of $5$~nm of Au on top of $5$~nm of Ni are {
{evaporated on the sample patterned by photolithography}}. 
The shape of the electrodes is sketched in Fig.~\ref{fig:fig2}~(b-c).  
The lateral dimensions are $l=10$~$\mu$m and  $L=145$~$\mu$m.
The electrodes ensure $\approx 1$~eV Schottky barrier shift with respect to bare surface \cite{Schmitz1996,Miura,Rickert2002,Chiaruttini2019} and are used in state-of-the-art GaN/(Al,Ga)N high electron mobility transistors \cite{Medjdoub2011}. 
The energy band alignment, calculated using NextNano++ software \cite{nextnano}, is affected by the electrodes as shown in   Fig.~\ref{fig:fig1}~(a). The black dashed line is calculated in the presence of the electrodes, while the solid line  is  calculated in their absence. 
The corresponding IX transition energies are shown in Fig.~\ref{fig:fig1}~(c).
The black star indicates the energy in the bare surface region and a black circle in the region covered by  the electrodes.
The energy gap between these two energies, $U_{t0}$, defines the depth of the trap in the absence of external bias and in the zero-density limit.
The resulting  zero-density in-plane potential at $V_b=0$ is shown in Fig.~\ref{fig:fig2}~(a)  by the black dashed line. 
The low-energy area, sandwiched between electrode-induced bumps of the potential, constitutes an electrostatic trap for IXs.
The same technology has been used in Ref.~\onlinecite{Chiaruttini2019} where electrostatic traps for GaN-hosted IXs have been demonstrated.

For application of an external electric bias, the outer part of the electrode surface is additionally covered by an opaque $200$~nm-thick gold layer to which the wires are bonded.
An electric bias $V_b$ between  $-3$~V and $1$~V is applied to the electrodes, while the naturally n-doped substrate ensures the common ground. 
The resulting current is measured in both  the absence and presence of optical excitation.
%
%
{Positive electric bias reduces the Schottky barrier, while a negative bias has the opposite effect.  Consequently, the energy bands shift as illustrated in Fig.~\ref{fig:fig1}~(a).  The red dashed line is calculated at positive bias, and the blue at a negative one. The corresponding IX energies are shown in Fig.~\ref{fig:fig1}~(c) with the same color code. 
The resulting profiles of the in-plane potential at positive, zero and negative biases are shown in Fig.~\ref{fig:fig2}~(a) by dashed lines.
One can see that the application of a negative bias further enhances the potential barriers  under the electrodes, while a positive bias smears them out.
This effect provides the basis for IX density control by an external bias.
}

Importantly, due to their mutual repulsion  IXs created in the trap screen out the trap potential $U_{t0}$  \cite{Lefebvre2004,Fedichkin2015,Fedichkin2016}. This screening results in a blueshift $E_{BS}$ of the IX energy in the trap:
\begin{equation}E_{IX}(n)=E_0 + E_{BS}(n),
  \label{eq:EBS_E0}
\end{equation}
and in the reduction of the trap depth:
\begin{equation} 
U_t(n)=U_{t0}-E_{BS}(n),
  \label{eq:Ut}
\end{equation}
as shown in Fig.~\ref{fig:fig2}~(a) by solid lines.
When excitonic correlation effects are negligible, $E_{BS}$ is usually assumed to be proportional to the IX density:  %
\begin{equation}
   E_{BS}(n)=\phi_0 n.
    \label{eq:EBS}
\end{equation} 
{For our sample  
$\phi_0=11.2\times10^{-11}$~meV$\cdot$cm$^{2}$. This value is extracted from} the self-consistent 
solution of the Schr\"{o}dinger and Poisson equations, see Appendix~\ref{sec:appendix_param} \cite{nextnano}. 
%
Equations (\ref{eq:EBS_E0}) and  (\ref{eq:EBS})   allow us to monitor the variations of the IX density in the sample.

Another important effect that needs to be accounted for is the exponential dependence of the electron and hole wavefunctions overlap on the IX density.
It leads to an exponential relation between IX emission intensity $I$ and its energy blueshift $E_{BS}$ \cite{Liu2016,MazuzHarpaz2017,Butov1999,Fedichkin2016,Chiaruttini2019}:
\begin{equation}
    I \propto E_{BS} \mathrm{exp} \left ( E_{BS}/\gamma \phi_0 \right ).
    \label{eq:IE}
\end{equation}
We estimate $\gamma=1.9 \times 10^{11}$~cm$^{-2}$ for our sample (see Appendix~\ref{sec:appendix_param}).

The sample is placed in a cold-finger cryostat and  cooled to $10$~K. Optical excitation is provided by a continuous wave laser focused in a $\approx 1$~$\mu$m-diameter spot on the sample surface, that we locate either within the trap on the bare surface, or on the semi-transparent electrode, see Fig.~\ref{fig:fig2}~(b) and (c), respectively. The excitation wavelength, $\lambda=355$~nm, is situated very close to the GaN bandgap energy, but well 
 below the emission energy of the Al$_{0.07}$Ga$_{0.93}$N barriers, to avoid spurious charge carriers in the structure. The microscope objective collects the PL signal in such a way that the image of the sample surface is reproduced at the entrance of a spectrometer, where it is filtered by a vertical entrance slit.  This setup provides a direct spatial imaging of the
PL with $\approx 0.8$~$\mu$m spatial and $\approx 1$~meV spectral  resolution.

\begin{figure}
	\includegraphics[width=3.4in]{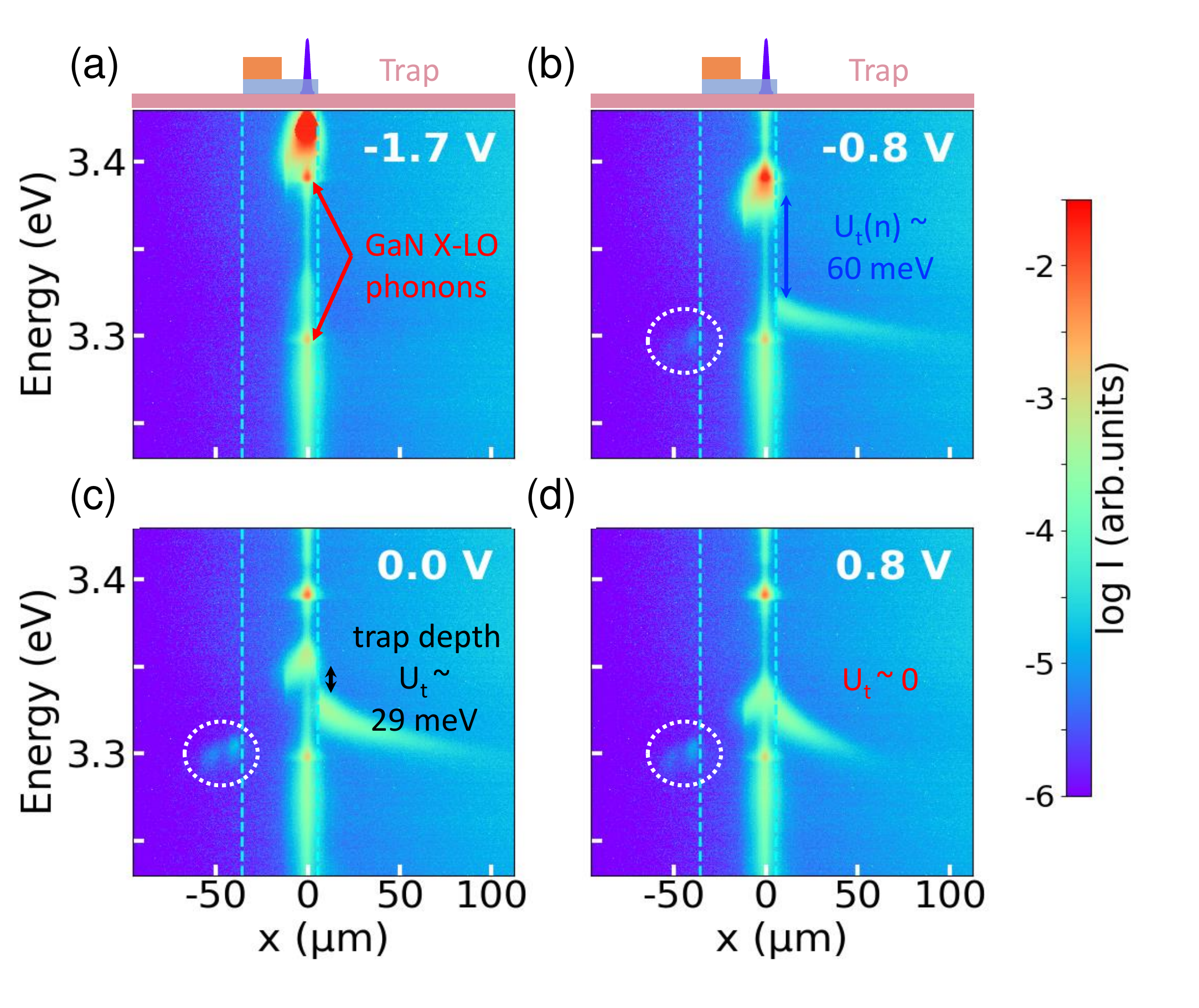}
	\caption{  Spatially resolved PL spectra (color-encoded in log scale) measured upon  excitation through the semi-transparent electrode at different values of the applied electric bias $V_b$. Both laser spot and electrode positions are indicated on top with the same color code as in Fig.~\ref{fig:fig1}~(b). Vertical dashed lines delimit electrode-covered area. Dashed circles highlight emission of IXs coming from outside of both  trap and  electrode-covered area. 
	}
		\label{fig:ColormapOutside}
\end{figure}


\section{Experimental results and discussion}
\label{sec:experiments}

The presentation of  experimental results is structured in two parts. 
In Section \ref{subsec:general} we  analyze how an applied electric bias modifies the in-plane potential pattern imposed by the electrodes. 
In the second part  we determine to which extent the IX density in the trap can be controlled by an applied electric bias (Section \ref{subsec:trap_eff}).

\subsection{General operation of the electrostatic trap}
\label{subsec:general}
In order to explore how the in-plane potential experienced by IXs is affected by the applied bias, we start from the measurements in the configuration shown in Fig.~\ref{fig:fig2}~(c): the excitation spot is located on top of the semi-transparent electrode. 
This choice of the excitation geometry allows us  to monitor the bias-induced potential variations via the energy of the IX emission passing trough the semi-transparent electrode. 

Fig.~\ref{fig:ColormapOutside} shows color-encoded (logarithmic scale) PL spectra measured at various distances from the spot along the x-axis. Measurements at four different values of the electric bias are shown, the incident power is $P=3.8$~mW. The entire set of  data comprising all the values of the applied bias is available as Supplemental Material. 

\begin{figure}
	\includegraphics[width=3.4in]{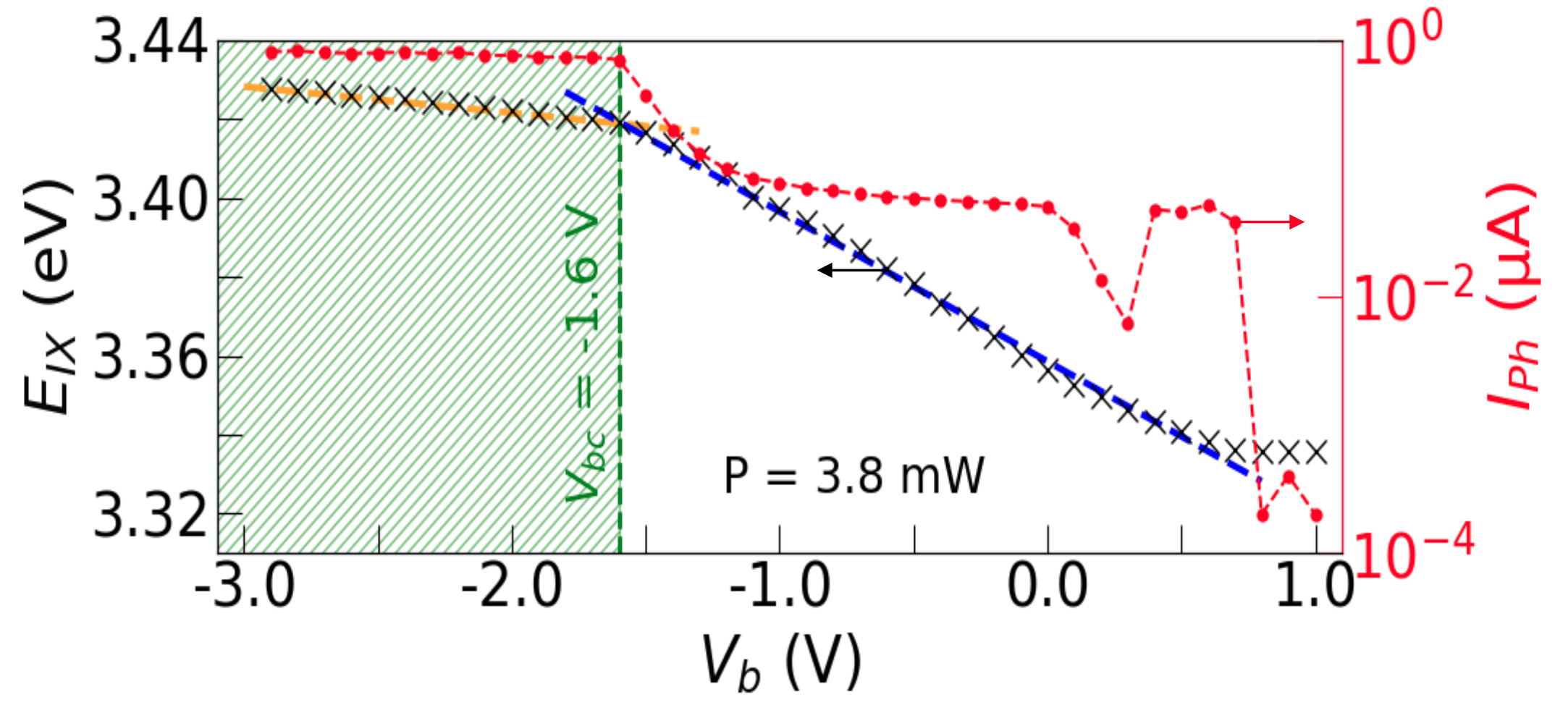}
	\caption{ Left axis: bias dependence of the IX emission energy  measured under the excitation spot ($x=0$) in the same geometry as in Fig.~\ref{fig:ColormapOutside} (crosses).  Dashed lines are linear fits  of $E_{IX}$ above (blue) and  below (orange) the critical voltage $V_{bc}$.  Right axis: bias dependence of the corresponding photo-induced current (red circles).  Hatched area shows the range of applied biases where photo-current saturates.
	}
		\label{fig:VbE}
\end{figure}

Let us first consider the measurements at $V_b=0$~V. 
Under the spot the emission is extremely broad. On top of this broad emission two very narrow peaks can be distinguished, at $3.390$~eV and at $3.298$~eV. They show up in all the spectra measured in this work, and can be attributed to
{one and two LO phonon replicas of the free exciton state in bulk GaN, respectively}. The broader peak at $\approx 3.355$~eV corresponds to  IXs emitting directly under the spot through the semi-transparent  electrode.
%

{Excitonic  emission can be observed on both sides of the excitation spot.} Its energy decreases continuously because the density of IXs decreases when they spread away from the excitation spot due to mutual repulsion. 
To the left from the spot, in the area covered not only by the thin, but also by the thick gold electrode, no optical emission is detected.
Further away, in the bare surface region at the border of the gold electrode, weak bias-independent IX emission at $\approx 3.3$~eV  can be distinguished. It is pointed out by white a dashed circle.

To the right from the spot, at $\approx 5-6$~$\mu$m,  diffusing IXs  reach the electrode edge and fall into the trap. Their  emission energy suddenly drops by  $\approx 29$~meV. This drop corresponds to $U_t(n)$, the in-plane potential barrier created by the electrodes and partly screened out by the trapped IXs, see Eq.~\ref{eq:Ut} and Fig.~\ref{fig:fig2}~(a). Further away the IX emission energy decreases continuously due to a reduction of the IX density via optical recombination and diffusion in the trap.

\begin{figure}
	\includegraphics[width=3.4in]{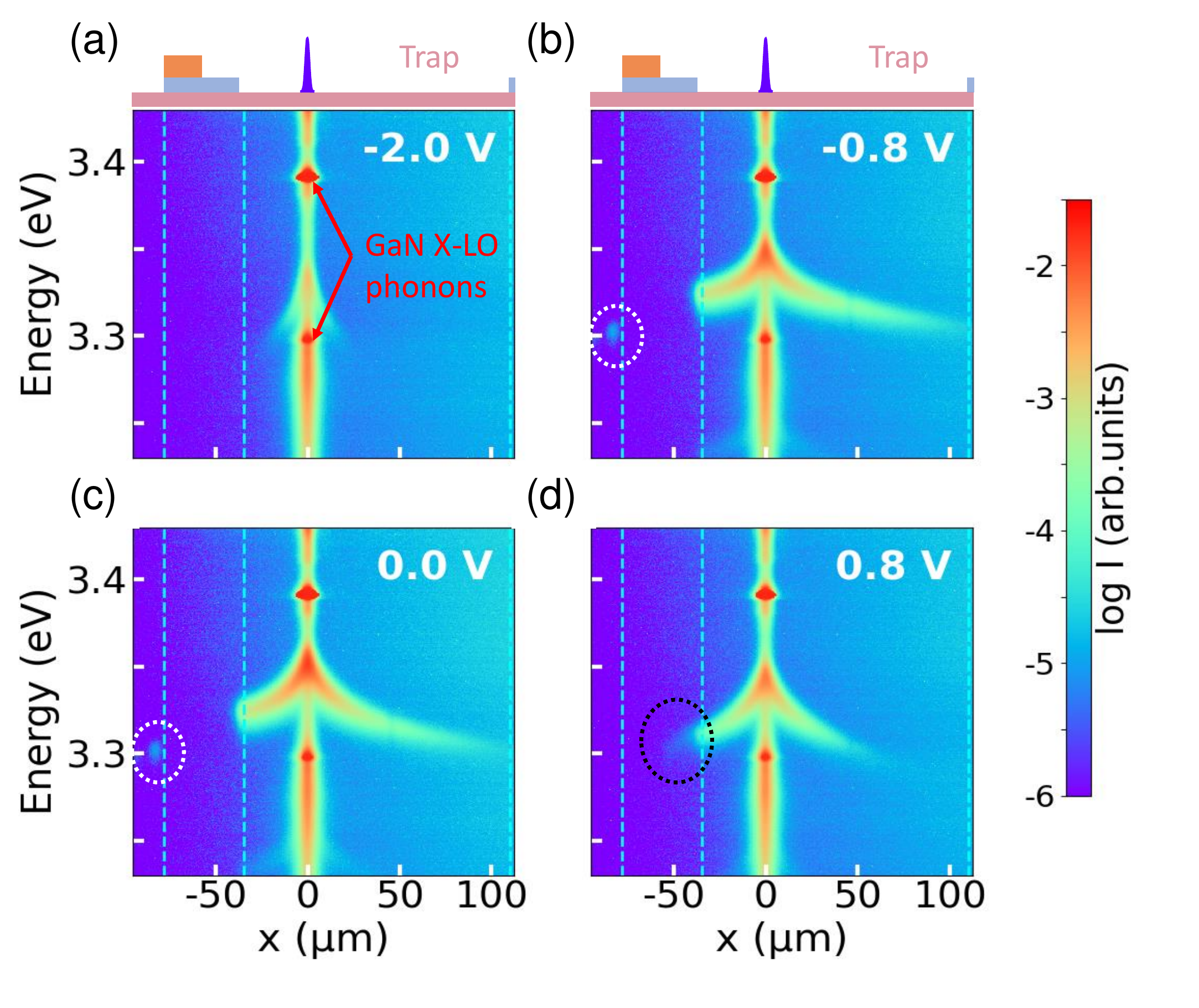}
	\caption{   Spatially resolved PL spectra (color-encoded in log scale) measured upon excitation at $x=0$ at different values of the applied electric bias $V_b$. Excitation power is $P=2.4$~mW. Excitation spot and electrode positions are indicated on the top with the same color code as in Fig.~\ref{fig:fig1}~(b). Vertical dashed lines delimit electrode-covered area. Dashed circles point IX emission  from outside of the trap (white) and from the area covered by the electrodes (black).
	}
		\label{fig:ColormapInsideHigh}
\end{figure}
\begin{figure}
	\includegraphics[width=3.4in]{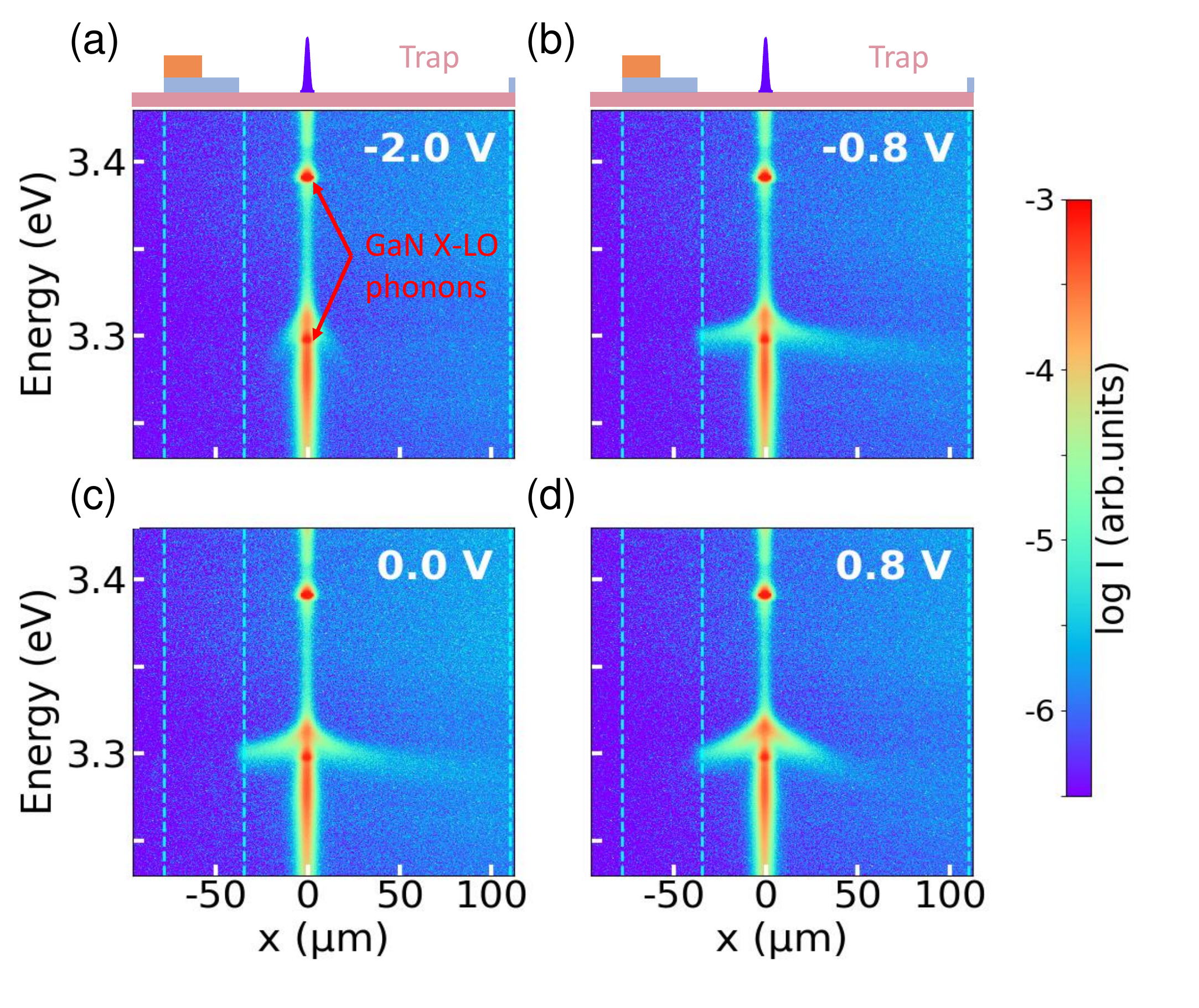}
	\caption{    Spatially resolved PL spectra (color-encoded in log scale) measured upon excitation at $x=0$ at different values of the applied electric bias $V_b$. Excitation power is $P=0.6$~mW. Excitation spot and electrode positions are indicated on the top with the same color code as in Fig.~\ref{fig:fig1}~(b).
	}
		\label{fig:ColormapInsideLow}
\end{figure}

Application of the positive bias $V_b=0.8$~V dramatically affects the emission pattern, see Fig.~\ref{fig:ColormapOutside}~(d). Indeed, the emission energy under the spot decreases, and the steep drop of the IX energy at the electrode border smears out. 
By contrast, application of the same, but negative bias $V_b=-0.8$~V (Fig.~\ref{fig:ColormapOutside}~(b)) produces the opposite effect. The emission energy under the spot increases and the energy drops stronger at the electrode edge, $U_t\approx 60$~meV. 
This is consistent with the idea presented schematically in Fig.~\ref{fig:fig2}~(a): a positive bias reduces the trap depth and thus the energy difference between IXs emitting under the electrodes and in the trap, while the negative bias affects the in-plane potential in the opposite way. 

Under negative bias higher than a critical value $V_{bc}=-1.6$~V, the PL of IXs is almost entirely quenched, see Fig.~\ref{fig:ColormapOutside}~(a).  We attribute this quenching to the in-plane component of the electric field that builds up at the electrodes edges  and gives rise to IX dissociation \cite{Hammack2006}. This mechanism of non-radiative losses will is discussed in more details in Sections~\ref{subsec:trap_eff} and \ref{sec:appendix_DD}.

The photo-current and the IX emission energy measured under the spot are shown in Fig.~\ref{fig:VbE} as a function of the applied bias. One can see that when changing the bias from positive to negative, the photo-current  $I_{ph}$ increases up to almost $1$~$\mu$A at  $V_{bc}=-1.6$~V, where it saturates. The increase of $I_{ph}$ is accompanied by the increase of the IX emission energy under the spot, that is in the electrode-covered region. 
{The blue dashed line in Fig.~\ref{fig:VbE} shows the linear fit of the IX energy dependence on the applied bias. The resulting slope, $\mathcal{R}_e=38$~meV/V, } can be compared with the simple estimation assuming that all the applied voltage drops linearly between the surface and GaN buffer situated $\mathcal{L}=d+2b=208$~nm below. This yields $\mathcal{R}_e=e d/\mathcal{L}=38$~meV/V, where $e$ is the absolute value of the electron charge. This is in  perfect agreement with the experimentally measured rate \footnote{More sophisticated full numerical calculation using Nextnano++ software \cite{nextnano} yields much smaller variation rate $\mathcal{R}_e=e d/\mathcal{L}=14$~meV/V, see Fig.~\ref{fig:fig1}~(c)}. 

By contrast, at strong negative bias, above $V_{bc}=-1.6$~V, where IX transport and emission from the trap are quenched, the energy variation rate  decreases abruptly down to $\mathcal{R}_e=6.7$~meV/V. 
{ The corresponding photo-current saturates (red symbols), while the dark current (not shown) continues to grow linearly with increasing bias. } This suggests that the majority of the photo-carriers is extracted from the system in the z-direction via the photo-current. These non-radiative losses also limit IX accumulation when they are directly created in the trap, as will be discussed in Section~\ref{subsec:trap_eff}. 

Thus, we have an experimental  demonstration that, at least at $V_b>V_{bc}$, an external bias effectively modifies the potential experienced by  IXs under the electrodes.  
Nevertheless, the following legitimate question  arises. Does the bias have any effect on the potential in the bare surface regions, that is within the trap? One can see in  Fig.~\ref{fig:ColormapOutside} that IX density varies substantially across the trap at a given bias  and is also bias-dependent. 
%
Therefore, it is not trivial to disentangle the  bias and the density effects on the resulting potential profile, and a neat solution should be found.

To do so we take advantage of the fact that under a given surface potential (or, equivalently, fixed value of the built-in electric field along the growth axis) the relation between IX emission energy $E_{IX}$ and its intensity $I$ is expected to obey  Eqs.~\ref{eq:EBS_E0} and \ref{eq:IE} for any IX density. 
In order to extend the range of accessible IX energies and intensities analyze the  experiments where laser excitation is situated inside the trap, see Fig.~\ref{fig:fig2}~(b). The corresponding spatially resolved PL emission maps at $4$ different values of $V_b$ are shown in Figs.~\ref{fig:ColormapInsideHigh} and  \ref{fig:ColormapInsideLow}. Note that to avoid undesirable effects of the laser-induced heating of excitons we limit our consideration to the spatial area situated at more than $10$~$\mu$m away from the spot.

\begin{figure}
	\includegraphics[width=3.4in]{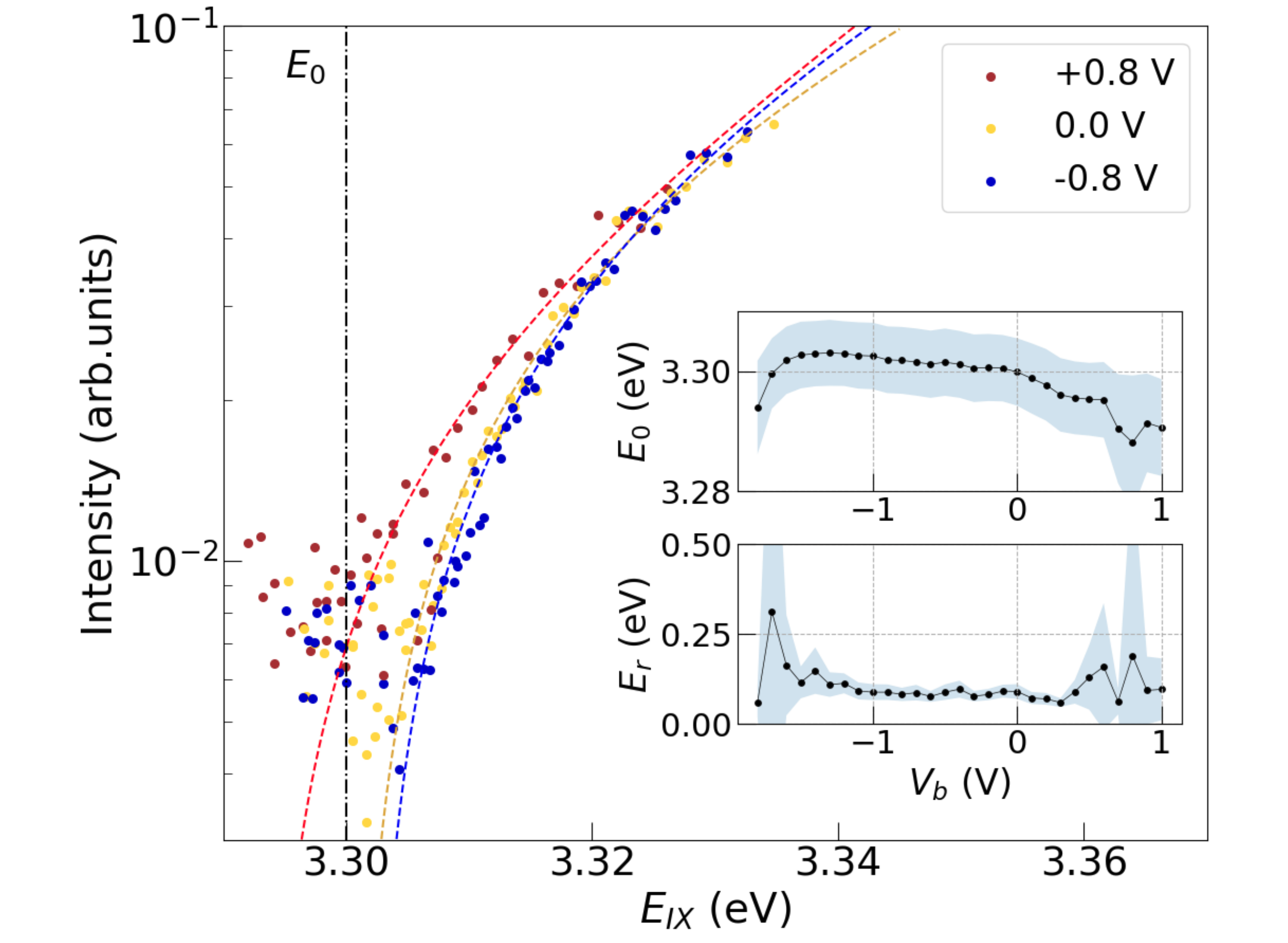}
			\caption{  Spectrally integrated intensity of the  IX emission as a function of its peak energy, extracted from the experiments as those shown in Figs.~\ref{fig:ColormapInsideHigh} - \ref{fig:ColormapInsideLow}  (excitation within the trap) at $V_b=0$ and at $V_b=\pm0.8$~V. Different energies correspond to different positions in the plane of the QW and we restrict our consideration to the points separated from the spot by  more than  $10$~$\mu$m. Dashed lines are fit to the data for each value of $V_b$ using Eqs.~\ref{eq:EBS_E0} and \ref{eq:IE}. Inset shows $E_0$ and $E_r = \gamma \phi_0$, obtained from the fitting procedure at different values of the applied bias.
    }	
		\label{fig:VbT}
\end{figure}

Fig.~\ref{fig:VbT} shows 
{ integrated IX emission intensities} measured at different positions within the trap as a function of the corresponding emission energies. 
Both energies and intensities are determined  by fitting the phenomenological function, introduced in Ref.~\onlinecite{Chiaruttini2021} and also reported in Appendix~\ref{sec:appendix_param}, to the IX emission spectra. 
The data obtained at $3$ different values of $V_b$ are presented.

The analysis of the dependencies shown in Fig.~\ref{fig:VbT} in terms of Eqs.~\ref{eq:EBS_E0}, \ref{eq:EBS} and \ref{eq:IE}  allows us to  quantify to which extent the electrostatic potential in the bare surface regions is affected by the applied bias,  and to reveal potential variations of $E_0$ and $E_r \equiv \phi_0 \gamma$ which are not expected from our simple modelling
\footnote{IX densities reached in these experiments remain below exciton Mott transition, consequently we neglect eventual density-dependence of the IX binding energy.}.
The results of the fitting procedure are shown by lines in Fig.~\ref{fig:VbT}. The summary of $E_0$ and $E_r$  values in the entire range of accessible bias voltages is given in  the inset.

One can see that  $E_r\approx 100$~meV remains constant within the error bars, while $E_0$, apart from the point at $V_b=-1.8$~V, seems to decrease  slightly when the bias is changed from zero to positive values \footnote{This behaviour could result from some spurious spread of the electric field from the metal-covered area towards bare surface regions.}.
%
%
However, the variation of $E_0$  remains much weaker than that of the trapping potential $U_{t0}$, suggesting that it could be neglected.
%

%
\begin{figure}[t]
	\includegraphics[width=3.4in]{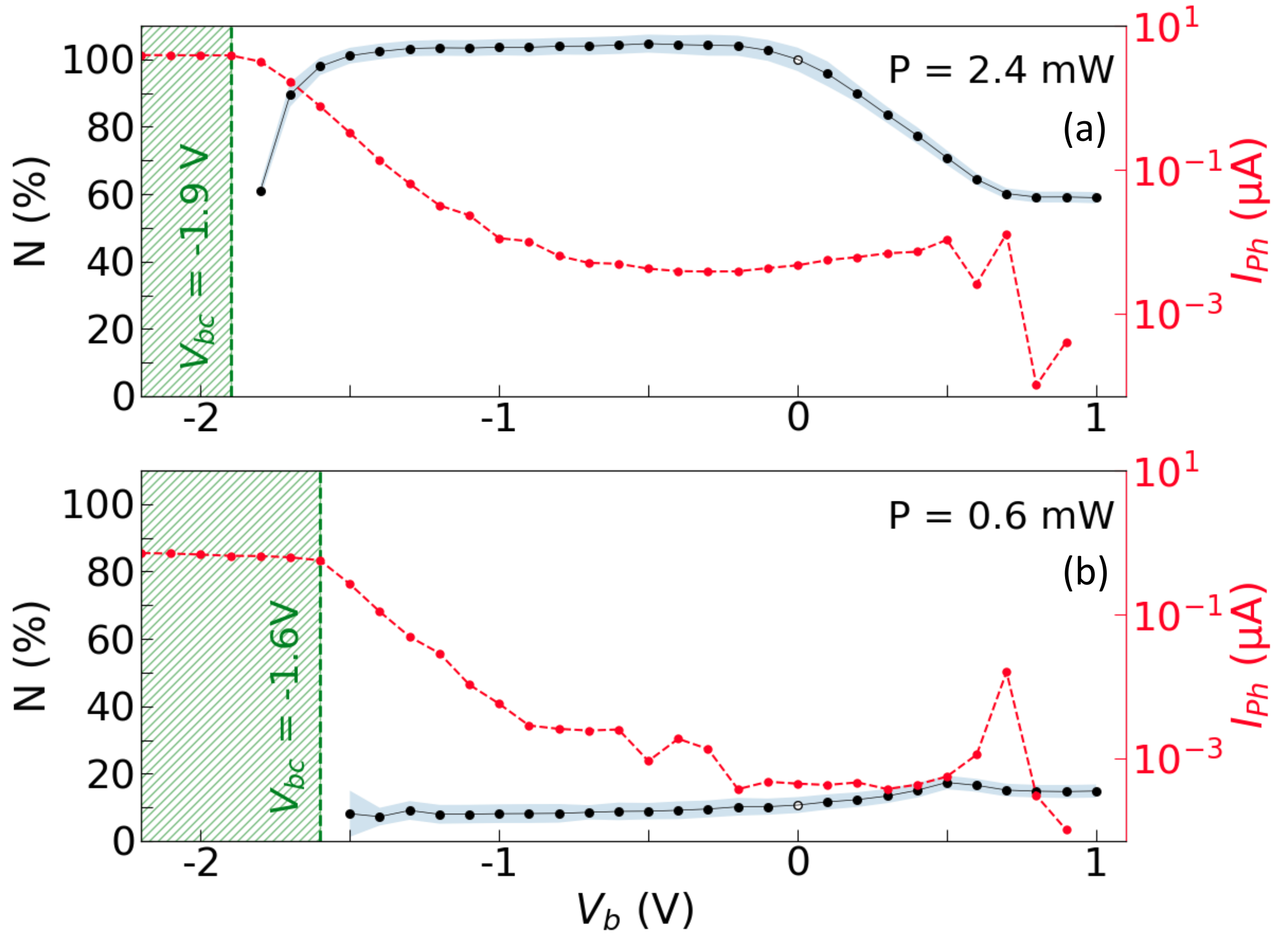}
	\caption{ Number of IXs within the detection area covering $\approx 20\%$ of the trap (reported in percents of $N_0$, IX population measured at excitation power $P=2.4$~mW and  $V_b=0$, left scale) and the corresponding 
	photo-current (right scale) as a function of the applied bias.  IXs are created directly in the trap either at $P=2.4$~mW (a) or at $P=0.6$~mW (b). Hatched area shows the range of applied biases where photo-current saturates.
	 Representative PL spectra corresponding to this analysis are shown in Figs.~\ref{fig:ColormapInsideHigh} and \ref{fig:ColormapInsideLow}.}
		\label{fig:DensityCurrent}
\end{figure}

It is instructive to notice, that at lowest emission energies -- mainly measured at largest distances from the excitation spot or at low excitation power -- the dependence of the IX emission intensity on its energy $I(E_{IX})$ starts deviating from the  exponential behaviour  typical for two-dimensional systems, {\it {cf}} Eq.~\ref{eq:IE}.
This points to the onset of the IX localization on the in-plane disorder potential at low IX densities.
Indeed, due to strong built-in electric field in the growth direction, the amplitude of the localizing potential  resulting from  both (Al,Ga)N alloy composition and QW width fluctuations can reach $10-20$~meV even in the best quality samples \cite{Weisbuch2021}.

In Appendix~\ref{sec:appendix_FWHM} we analyse the relation between the IX emission energy and its linewidth (defined here as full width at half maximum (FWHM) of the  exciton emission line) and unravel homogeneous and inhomogeneous contributions to the IX linewidth.
This analysis corroborates the idea that fundamental properties of IXs inside the trap remain  unaffected  by 
%
{the external bias}.   
Thus, weak variations of $E_0$ and $E_r$  will not be accounted for in the estimations of the trapped IX densities reported below. We conclude that  $E_0=3.300\pm0.006$~eV for any $V_b$, while the trapping potential varies in the presence of the external bias at 
{the rate $\mathrm{R}_e\approx 38$~meV/V}. 

\subsection{Bias-controlled exciton trapping and release}
\label{subsec:trap_eff}
%
%
%
%
Let us now address the efficiency of the IX density control by the external bias  and identify the conditions that maximize and minimize the density of  IXs accumulating in the trap.

%
For this purpose  we analyze the emission of  IXs created directly within the trap at two different powers, $P=2.4$~mW and $P=0.6$~mW. The corresponding PL maps are shown in Figs~\ref{fig:ColormapInsideHigh} and~\ref{fig:ColormapInsideLow},  respectively. The same set of data was used in the the analysis reported in Fig.~\ref{fig:VbT}. The entire set of the data
corresponding to all the values of the applied bias is
available as Supplemental Material.  

%
%

The main features that we observe are the following: (i) at zero applied bias, as well as  at $V_b=-0.8$~V, IXs accumulate at the 
{ left-hand trap edge}. This is evidenced by the spatial asymmetry of the emission energy (or, equivalently, IX density) profile.
At higher excitation power a fraction of these accumulated IXs escapes from the trap and emits light outside the electrode-covered area. Their energy is close to the zero-density exciton energy $E_0=3.3$~eV. This emission is pointed out by white dashed circles.
(ii) At high negative bias $V_b=-2$~V the emission is strongly quenched, in the same way as in the case of excitation through the electrode, {\it cf} Fig.~\ref{fig:ColormapOutside}~(a). 
%
(iii) At positive bias  $V_b=0.8$~V the IX density distribution around the excitation spot is much more symmetric than at zero bias.  This is consistent with the reduction of the trap depth at positive bias that is also observed in Fig.~\ref{fig:ColormapOutside}~(d).


To quantify the efficiency of the IX trapping we show in Fig.~\ref{fig:DensityCurrent} the total number of IXs within the detection area in the trap ($2$~$\mu$m $\times$ $145$~$\mu$m, $\approx 20\%$ of the total trap area) 
as a function of the applied bias.  For both excitation powers IX populations are given in percents of $N_0$, the number of IXs measured at $V_b=0$ and $P=2.4$~mW. Using Eq.~\ref{eq:EBS_E0} with $E_0=3.3$~eV  it can be estimated as $N_0 \approx 2.6\times 10^{13}$ particles.
Note that only IXs emitting at energies above $E_0$, which are supposed to be delocalized, are taken into account (see Sec.~\ref{subsec:general} and Appendix.~\ref{sec:appendix_FWHM}).



One can see in Fig.~\ref{fig:DensityCurrent} that the population of IXs in the trap does not grow up at negative bias. Moreover, above a critical bias $V_{bc}$  IX population decreases sharply,  in concert with the saturation of the photo-current. The latter is shown on the right-hand scale of Fig.~\ref{fig:DensityCurrent}. As  mentioned in Sec.~\ref{subsec:general} 
{and sketched in Fig.~\ref{fig:fig2}~(a)}, we explain this effect as being due to IX dissociation induced by the in-plane component of the electric field $F_{xy}$ that builds up at the  electrode edges.
%
The interpretation of this phenomenon within drift-diffusion model coupled to bias-dependent non-radiative losses allows us to reproduce the experimental results at least qualitatively, see Appendix~\ref{sec:appendix_DD}.

Importantly, the critical bias is density-dependent, it decreases with the excitation power: $V_{bc}=-1.9$~V at $P=2.4$~mW and $V_{bc}=-1.6$~V at $P=0.6$~mW.
This can be understood as follows.
At high power IXs accumulated in the trap partly screen out the trap potential $U_{t0}$  imprinted by the electrodes, so that  $U_t(n)<U_{t0}$. As a result, the trap depth is reduced by $E_{BS}$, see Eq.~\ref{eq:Ut} and  Fig.~\ref{fig:fig2}~(a). Therefore, at a given bias the in-plane field $F_{xy}$ leading to exciton dissociation and losses via photo-current is reduced as well. 
Thus, the critical value of $F_{xy}$  such that  the photo-current dominates over the radiative recombination in the decay of the IX population  is reached at a higher negative bias in the strongly excited trap.

Another effect of the pumping power on the trapping efficiency manifests itself at positive bias. Indeed, as one can see in Fig.~\ref{fig:DensityCurrent}~(a), at high power IXs are efficiently released from the trap when a positive bias is applied, their population is reduced by $\approx 40\%$. In contrast, at low power (Fig.~\ref{fig:DensityCurrent}~(b)) IX population remains constant.
We argue  that this is because, at low densities, close to the localization threshold, IX diffusion is not efficient enough to allow them to escape from the trap.  
%
%
%
%
%

{Altogether, while we could not alter IX population in the trap by applying a negative bias, application of a positive bias allowed us to unloads IXs from the trap, here up to $40\%$.} The implementation of such control appears to depend crucially on the trap filling because it relies on efficient diffusion, which is density-activated \cite{Fedichkin2016}. 
The optimum trapping conditions, characterized by maximum IX density collected within the trap, are reached at zero applied bias. This  differs advantageously from GaAs and TMD-based systems hosting IXs, where application of the bias is a prerequisite for IX trapping.

\section{Summary and Conclusions}
\label{sec:conclusions}
In conclusion, we have studied the effect of an external electric bias on the operation of the electrostatic trap for IXs. 
Such device is formed by metallic gates deposited on the surface of polar GaN/(AlGa)N QW. 
The particularity of this type of QWs consists in a strong built-in electric field, typically several hundreds of volts per centimeter.
It  has been demonstrated earlier that deposition of metallic gate reduces this field, so that in contrast with their GaAs-hosted counterparts, IXs in GaN/(AlGa)N QWs get trapped in the regions of the QW plane that are not covered by electrodes. 

In this work we have shown that the trapping potential can be additionally controlled   by the external electric bias.
In a similar way as in GaAs-based devices developed for IX trapping, when a bias is applied between the n-doped substrate and the metallic gate,  IX energy levels in the QW are pushed either up or down, depending on the sign of the applied bias. 
The resulting variation of the in-plane trapping potential of order of several tens of millielectronvolts  has been evidenced by spatially-resolved $\mu$PL spectroscopy across semi-transparent electrodes.
These experiments also demonstrate that IX energy in the bare surface regions is only weakly affected by the bias.

The accumulation of IXs in the trap at negative voltages, where the trap gets deeper, appears to be inhibited by the non-radiative losses resulting from the in-plane component of the electric field in the vicinity of the electrode edges.
These losses are detected as a photo-current that grows exponentially with increasing negative bias.
Above the IX density-dependent critical value of negative bias these non-radiative losses overcome radiative recombination. At this point the excitonic PL suddenly drops, and the photo-current saturates. 
Thus, at least within the current device geometry it is not possible to 
increase the IX density in the trap by applying an external electric bias. 
The modifications in the device design that we anticipate to  overcome this problem and reduce the in-plane field include the positioning of the QW closer to the homogeneous  bottom n-type layer, as suggested in Ref.~\onlinecite{Hammack2006}.

Application of positive voltages, that reduce both the trap depth and spurious in-plane electric field, allows us to release IXs out of the trap. Here the limitation is only imposed by the IX density, that needs to be sufficiently high.
Indeed, at low IX densities the disorder-driven localization hinders IX diffusion. Consequently, IXs can't move away from the trap during their lifetime and their  population inside the trap does not change despite the reduced trap depth.
Working with smaller traps could extend the density range where IXs release from the trap remains effective.

Finally, since IX diffusion in GaN/(AlGa)N QWs over tens of micrometers can be monitored optically, future work in the optimized devices may include room temperature IX density control by the external bias. This paves the way towards IX-based opto-electronic devices. 
On the other hand, provided that non-radiative losses related to electrode edges could be reduced, more complex multi-electrode devices, inspired by GaAs-based technology, but taking advantage of GaN particularities, may be interesting to address in the future.  

\begin{acknowledgments}
This work was supported by French National Research Agency via IXTASE (ANR-20-CE30-0032) and LABEX GANEXT projects.
\end{acknowledgments}

\section{Appendix}
\setcounter{equation}{0}
\label{sec:appendix_param}

\subsection{Density-dependence of IX energy and radiative recombination rate }

\renewcommand{\theequation}{X\arabic{equation}}
\begin{figure}
	\includegraphics[width=3.4in]{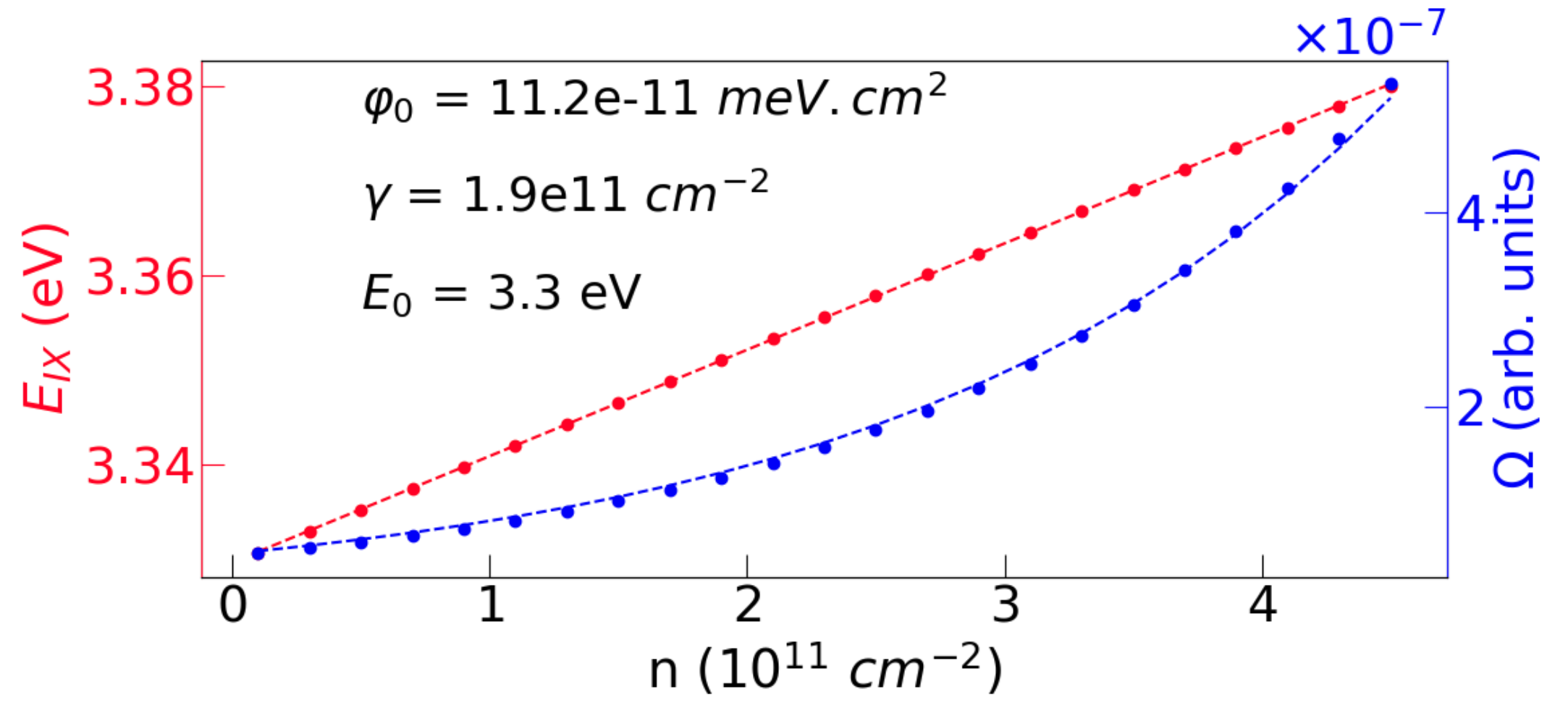}
	\caption{ Electron-hole pair energy (red) and squared overlap integral (blue)  calculated using NextNano++ software  as a function of the carrier density. }
		\label{fig:phi0}
\end{figure}
Fig.~\ref{fig:phi0} shows density dependence of the electron-hole pair energy (red symbols) and of the squared overlap integral (blue symbols) calculated for the sample under study using NextNano++ software \cite{nextnano}. 
The pair energy varies linearly with the density. The corresponding slope is estimated as $\phi_0=11.2$~meV/$10^{11}$cm$^{-2}$, see red dashed line. We assume that at not-too-high densities addressed in this work the IX binding energy is almost density-independent  so that the variation of IX energy with density is also linear and given by the same parameter: $E_{IX}=E_0+\Delta_{BS}=E_0+\phi_0 n$.
The calculated density dependence of the squared overlap integral $\Omega(n)$, and thus of the radiative lifetime of the electron-hole pair $\tau_R(n)$, exhibits exponential behavior: $\Omega\propto\tau_R \propto -n/\gamma$, see the blue dashed line \cite{Lefebvre2004,Fedichkin2015}. Fitting procedure yields $\gamma=1.9\times10^{11}$~cm$^{-2}$.  
Since the integrated IX emission intensity is given by $I=n/\tau_R(n)$, the relation between the intensity and the energy takes the form given by Eq.~\ref{eq:IE}. 
Note that in addition to one-dimensional calculations reproducing infinite-plane hetero-structure in the z-direction, NextNano++   offers the opportunity to calculate the electronic structure of the entire sample, including the whole heterostructure along the growth direction, and the top electrodes defining the electrostatic trap. Such calculations are needed to estimate the in-plane component of the electric field $F_{xy}$ acting on the  electron and hole constituents of the IX in the vicinity of the electrode edges. This field governs the non-radiative recombination of IXs that  needs to be accounted for in the model presented in Appendix~\ref{sec:appendix_DD}.

\subsection{Density-dependence of the IX linewidth }
\label{sec:appendix_FWHM}

The determination of exciton densities has always been a thorny issue. In this work we tentatively relay on a combination of two facts. (i) The IX emission energy shift $\Delta_{BS}=E_{IX}-E_0$  increases linearly with the density, at least in the absence of the excitonic correlation effects \cite{Remeika2009,Laikhtman2009,ZimmermannComment2010,Schinner2013,Remeika2015,Beian2017}  (ii) The relation between the IX density and intensity can be used to determine the zero-density energy $E_0$, so that $n=\Delta_{BS}/\phi_0$.
In the materials with high localization energies, the zero-density energy $E_0$ ought to be interpreted as the lowest energy below which excitons are essentially localized.

\begin{figure}
	\includegraphics[width=3.4in]{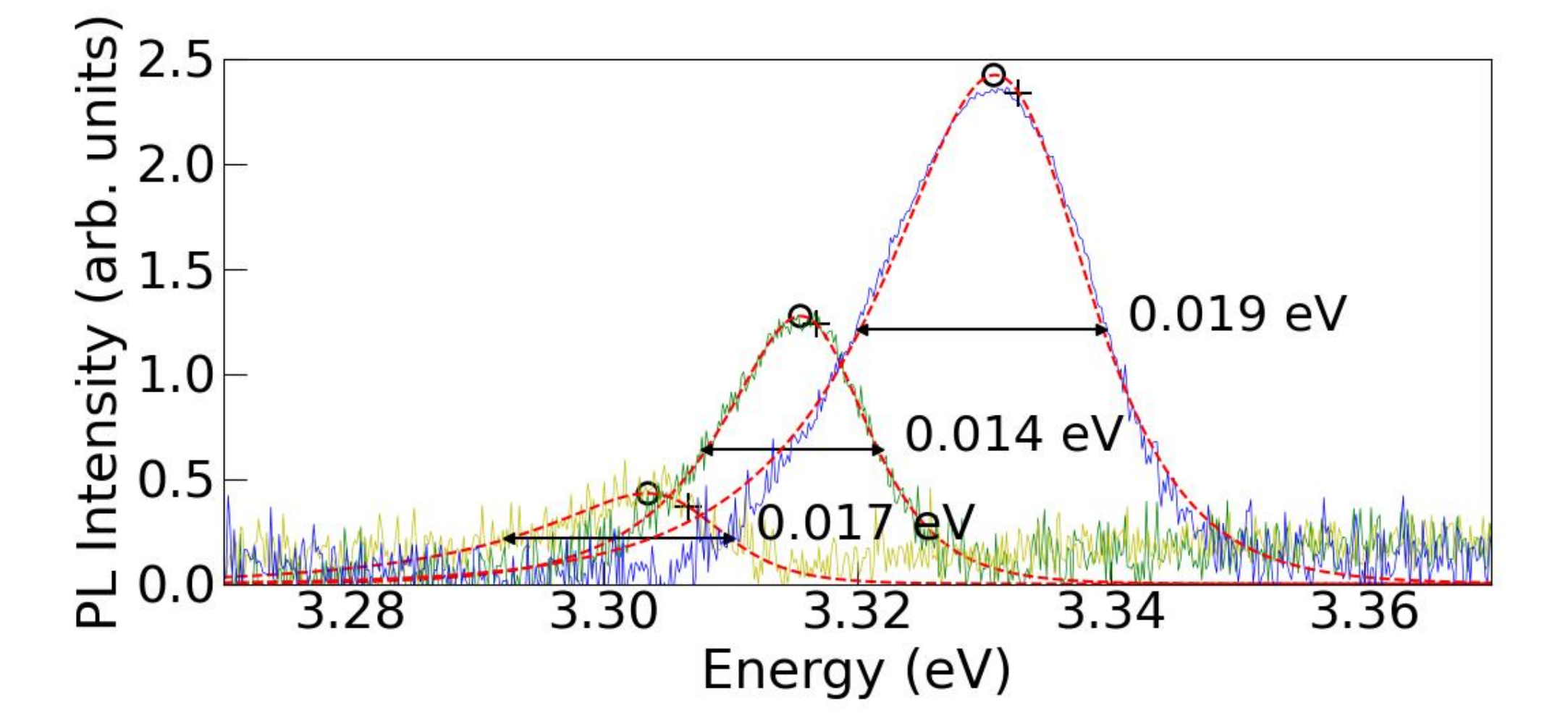}
	\caption{ PL spectra  measured at zero bias at $3$ different distances from the spot (blue: 15 µm, green: 45 µm, yellow: 95 µm), extracted from Fig.~\ref{fig:ColormapInsideHigh}~(c). Red dashed lines show the fits to Eq.~\ref{eq:fit}. For each spectrum, values of $E^{'}_{IX}$ (crosses) and the PL peak maximum $E_{IX}$ (circles) are indicated.}
		\label{fig:spectra}
\end{figure}

To further confirm and complement the determination of $E_0$ from the analysis of the IX emission intensity reported in Section~\ref{subsec:general},  we consider here the linewidth of the IX emission. The latter appears to be strongly density-dependent. 
Typical PL spectra measured within the trap at 3 different distances from the excitation spot at $V_b=0$ are shown in Fig.~\ref{fig:spectra}. 
Remarkably, the dependence of the linewidth on the emission energy (and thus on the  density) is non-monotonous. 
We suggest that this is due to interplay between homogeneous $\Gamma_h$ and inhomogeneous $\Gamma_i$ components of the linewidth \cite{High2009}
\begin{equation}
FWHM=\sqrt { \Gamma_h^2 +\Gamma_i^2}, 
\label{eq:gammatot}
\end{equation}
that evince opposite density dependencies. 
Indeed, at low density the disorder-induced inhomogeneous broadening  dominates the linewidth. The increasing density tends to screen out the disorder, leading to the reduction of the IX linewidth down to some density-independent value \cite{Ivanov2004,Savona2007,High2009}.
We model this effect by the phenomenological function:
\begin{equation}
    \Gamma_i=\Gamma_{i0}(1+\mathrm{exp}(-\Delta_{BS}/\sigma_i)).
    \label{eq:gammainh}
\end{equation}

In contrast, as far as IXs are localized, $E_{IX}<E_0$, the homogeneous contribution to the linewidth $\Gamma_{h0}$ is close to zero, $\Gamma_{h0}<0.1$~meV. It is expected to increase at  $E_{IX}>E_0$, and eventually to take over the inhomogeneous contribution. This is due to collisional broadening, which increases linearly with IX density  \cite{Honold1989,Voros2009,Gribakin2021}:
\begin{equation}
\Gamma_h=\Gamma_{h0}+\sigma_h \Delta_{BS}.
   \label{eq:gammahom}
\end{equation}
Here we define $\Delta_{BS}\equiv 0$ for $E_{IX}<E_0$.

\begin{figure}
	\includegraphics[width=3.4in]{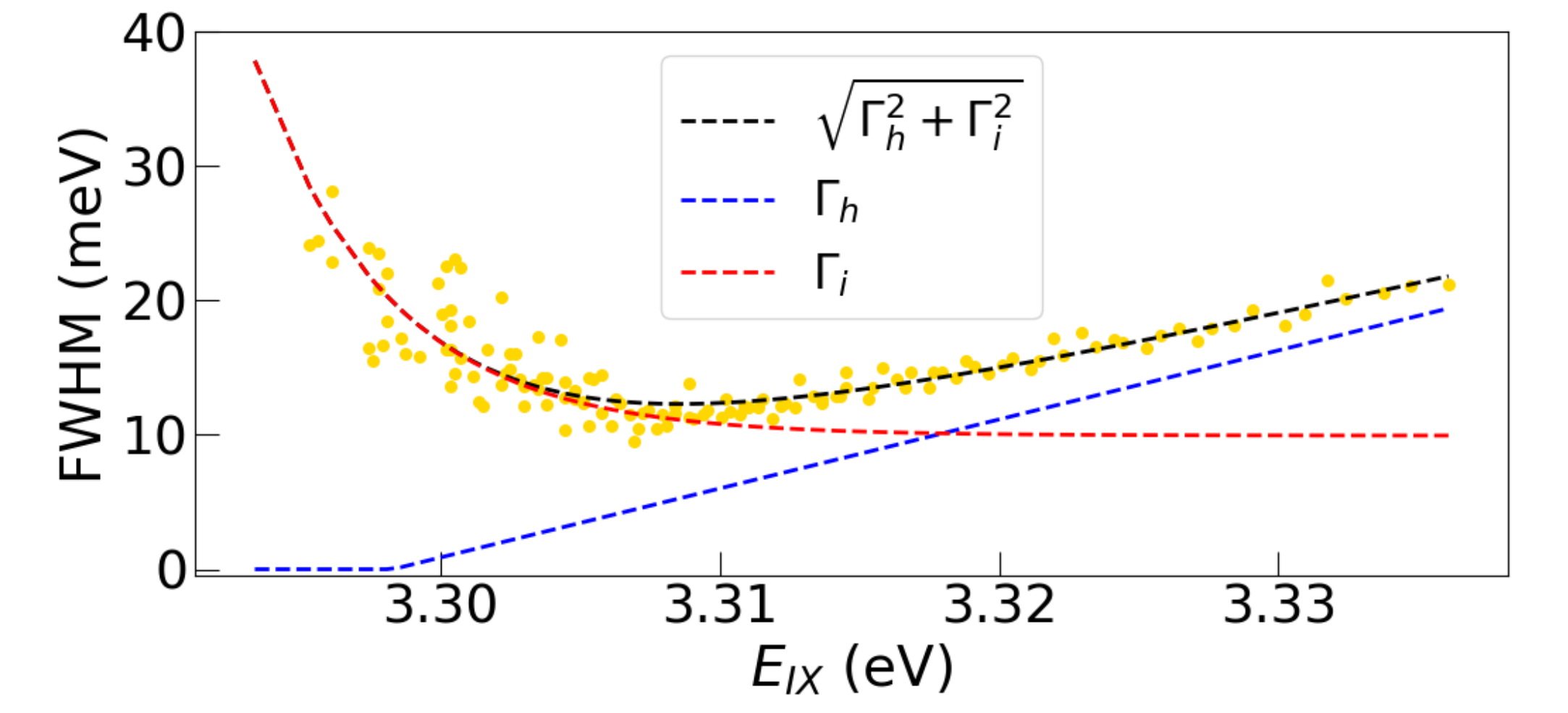}
	\caption{ Dependence of the IX linewidth (symbols) on its energy extracted from spatially-resolved PL at zero applied bias shown in  Fig.~\ref{fig:ColormapInsideHigh}~(c), and the corresponding fit to Eq.~\ref{eq:gammatot} (black line). Blue and red lines show homogeneous  and inhomogeneous  contributions to the FWHM, respectively. }
		\label{fig:linewidth}
\end{figure}
\begin{figure}
	\includegraphics[width=3.4in]{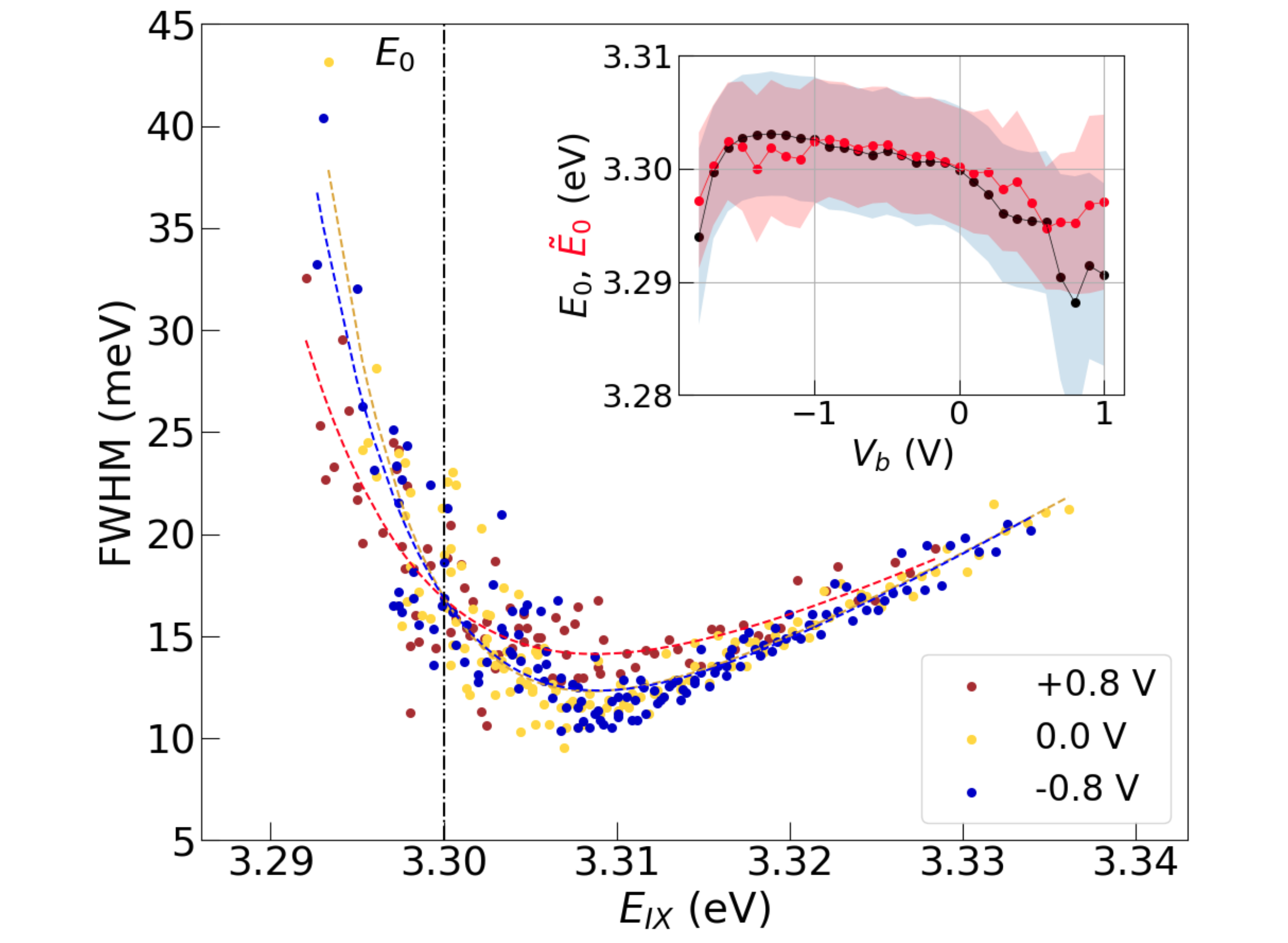}
	\caption{ Circles: IX linewidth  as a function of its energy extracted from spatially-resolved PL shown in  Fig.~\ref{fig:ColormapInsideHigh}  and \ref{fig:ColormapInsideLow}~(b-d). Dashed lines: fit to  Eq.~\ref{eq:gammatot}. Vertical  line shows the average zero-density value $E_0$ used for the analysis of the trapping efficiency in Section~\ref{subsec:trap_eff}. Inset: Parameter $\tilde{E}_0$ (red) extracted from this fitting procedure and  $E_0$ extracted from $I(E_{IX})$ analysis reported in the main text (black) as a function of the applied bias.}
		\label{fig:linewidth3Vg}
\end{figure}

The simple model described above allows us to analyze the linewidth dependence on the energy, and extract relevant parameters, in particular $E_0$ and its eventual bias dependence.
We proceed as follows. 
First of all we determine the IX properties -- emission energy $E_{IX}$, integrated intensity $I_{IX}$ and FWHM -- by fitting the function: 
\begin{equation}
I_{PL}(E)\propto \frac{ \mathrm{exp} ({\Delta}^{'}_{BS}/\beta_1) }
{1+\mathrm{exp}\left({\Delta}^{'}_{BS}/\beta_2\right)}
\label{eq:fit}
\end{equation}
to each spectrum. 
Here $\beta_1$, $\beta_2$ and $\Delta^{'}_{BS}$ are fitting parameters. Note, that the energy ${E}^{'}_{IX}={E}_{0}+\Delta^{'}_{BS}$ does not coincide with the maximum of the PL intensity $E_IX$, see Fig.~\ref{fig:spectra}. 
%

For each value of the applied bias, the dependence of the FWHM on $E_{IX}$ can thus be analyzed. A representative example of such analysis at $V_b=0$ is shown in Fig.~\ref{fig:linewidth}. Both homogeneous (blue) and inhomogeneous (red) contributions to the line broadening obtained from fitting of Eqs.~\ref{eq:gammatot}-\ref{eq:gammahom} to the data are shown. This offers an alternative way to determine the zero-density  IX energy, that we refer to as $\tilde{E}_0$. 
At $V_b=0$  we obtain $\tilde{E}_0=3.298$~eV, very close ${E}_0=3.300$~eV.
According to this analysis, the inhomogenous broadening dominates IX linewidth up to almost $20$~meV above zero-density energy. 
Fig.~\ref{fig:linewidth3Vg} shows the dependence of IX linewidth on its energy for three values of $V_b$, as well as the corresponding fits to Eqs.~\ref{eq:gammatot}-\ref{eq:gammahom}. The inset summarises the resulting values of $\tilde{E}_0$ (red), and compare them to  zero-density energy ${E}_0$ (black) extracted from the integrated intensity data.  
It appears that the two methods used to determine zero-density energy  yield quite similar results, corroborating its negligible dependence on the applied bias. 
Therefore, for tentative estimation of the IX densities shown in Figs.~\ref{fig:DensityCurrent} and~\ref{fig:freefem} we use a constant energy $E_0=3.3$~eV for all the values of applied bias.

Finally, it is worth noting that the parameters of the homogeneous broadening  $\sigma_h\approx 0.5\pm 0.1$ (see Eq.~\ref{eq:gammahom}), as well as of the inhomogeneous broadening  $\Gamma_{i0}=10\pm 2$~meV  and $\sigma_{i}=0.005\pm 0.002$ (see Eq.~\ref{eq:gammainh}),  that we  deduce from the fitting procedure do not show any  bias dependence. This may indicate that collisional broadening of IXs in the trap is governed by either exciton-exciton or exciton-impurity, rather that exciton-electron collisions.
One can compare $\sigma_h$ with the simplest theoretical estimation  $\sigma_h=e \phi_0 D_s$, where $D_s=M_x/(2\pi\hbar^2)$ is the two-dimensional density of states per spin, $\hbar$ is reduced Planck constant, and $M_x$ is the in-plane mass of the exciton \cite{Remeika2015}.
This yields $\sigma_h\approx 45$, almost an order of magnitude above the experimentally determined value. Similar results are reported in  Ref.~\onlinecite{Remeika2015} for IXs hosted by GaAs-based heterostructures.

\subsection{Drift-diffusion-losses modeling of the bias-controlled IX trapping }
\label{sec:appendix_DD}
\begin{figure}
	\includegraphics[width=3.4in]{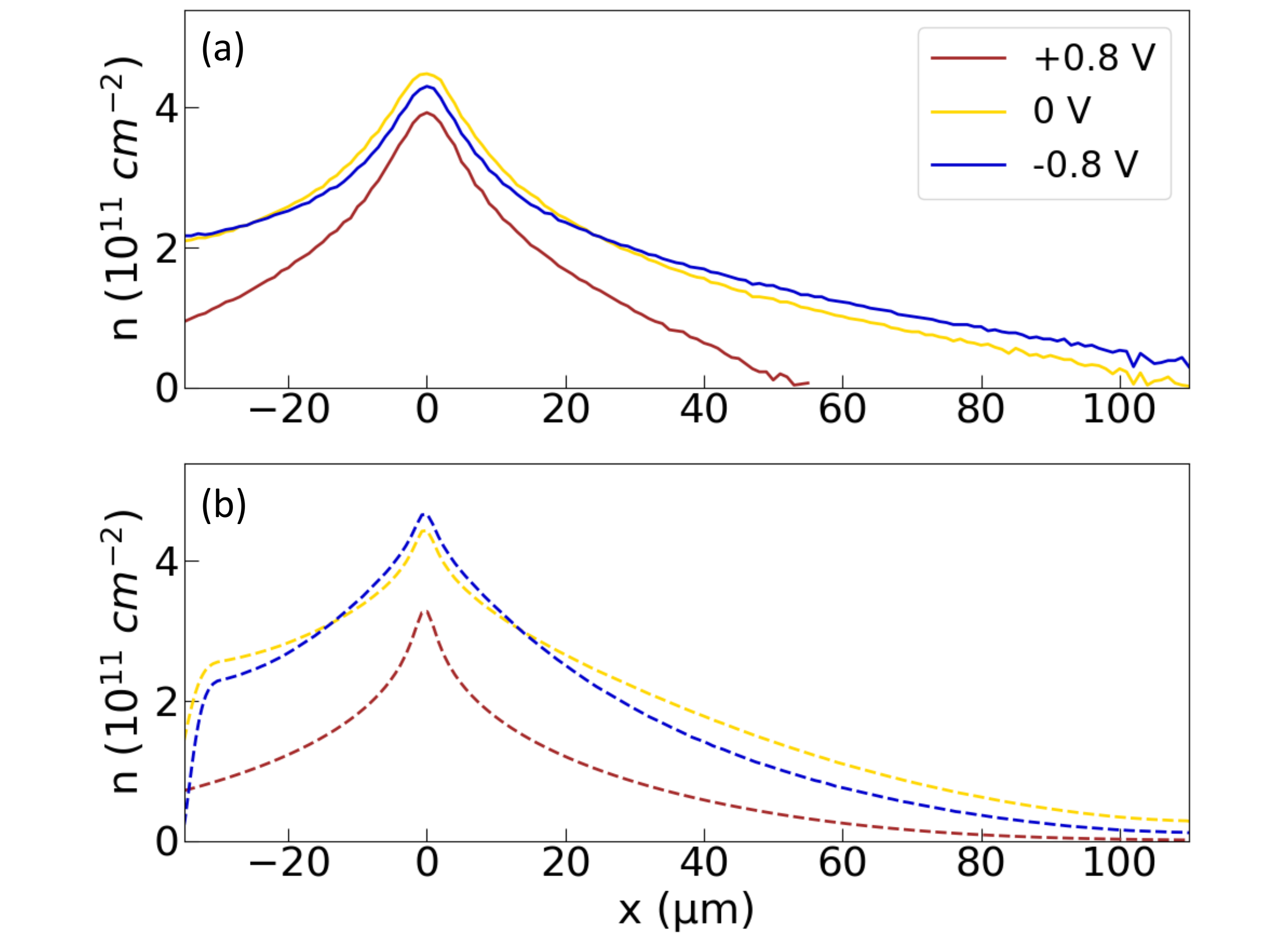}
	\caption{Distribution of IX density within the trap $n(x, y=0)$ measured (a) and calculated (b) for three different values of electric bias. Excitation spot is positioned at $x=0$. Experimental values of the IX density correspond to the data shown in Fig.~\ref{fig:ColormapInsideHigh}~(b)-(d).}
		\label{fig:freefem}
\end{figure}
\begin{figure}
	\includegraphics[width=3.4in]{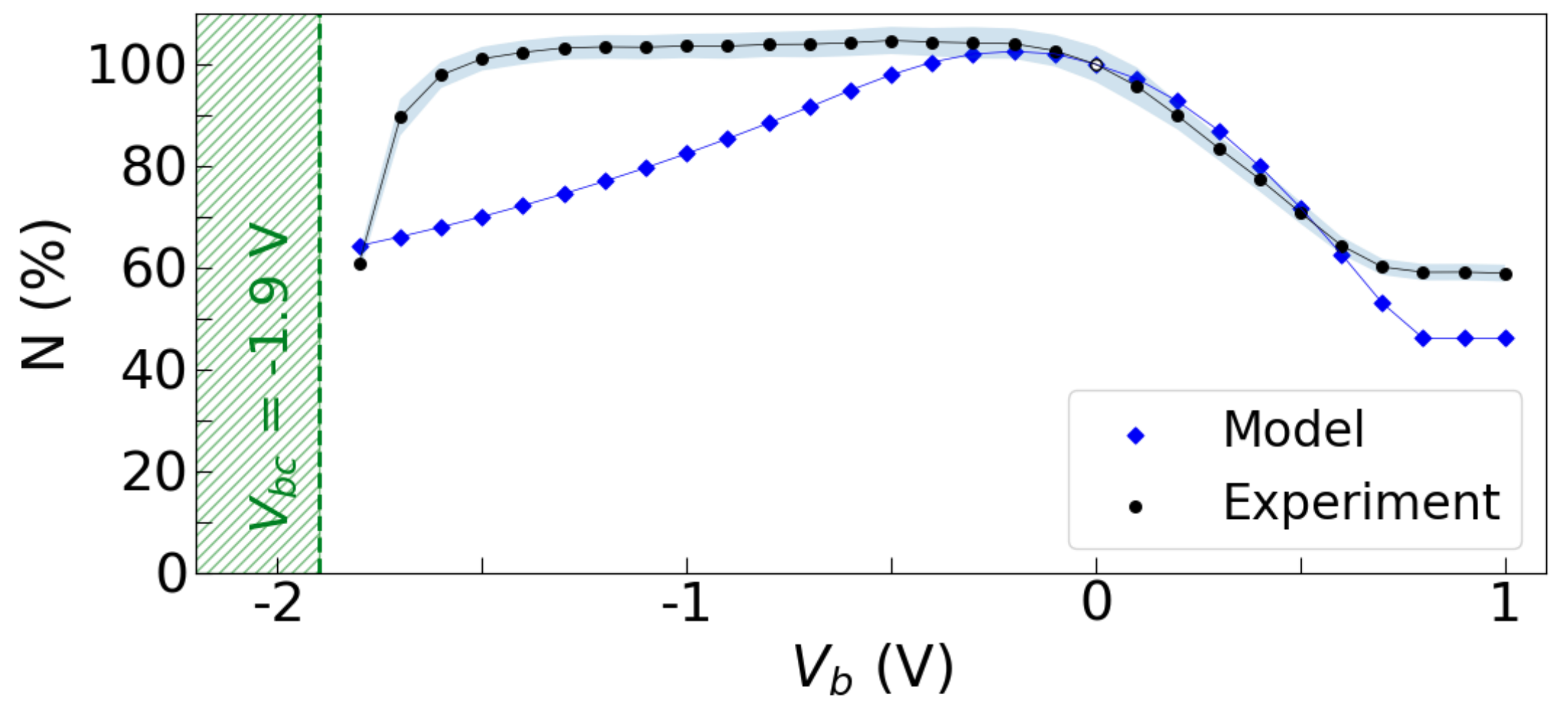}
	\caption{ IX population in the trap at $y=0$ integrated along x-axis (in percents of IX population at $V_b=0$), measured (red) and calculated (black) as a function of applied bias. Experimental values are the same as in Fig.~\ref{fig:DensityCurrent}~(a).
}
		\label{fig:freefemAllVb}
\end{figure}
The bias-dependent IX trapping can be modelled within two-dimensional drift-diffusion-losses formalism.
It is similar to the one used in our previous work, but includes two additional ingredients, namely bias-dependencies of the in-plane potential  and of the non-radiative recombination rate.

 The equation for the IX density under optical pumping reads:\cite{Ivanov2002,Rapaport2006}
\begin{equation}
\frac{\partial n}{\partial t}=
- \nabla \cdot \mathbf{J} + G - R n,
\label{eq:drift-diffu}
\end{equation}
where $R$ is the recombination rate, $\mathbf{J}$ is the IX current density.
The IX generation rate is $G= \left(N_p / \pi a_0^2 \right)  \exp (-(x^2+y^2)/a_0^2)$,
where $N_p$ is the number of excitons generated per second and $a_0=1$~$\mu$m is the radius of the excitation spot centered at $x=y=0$.  

The exciton current $\mathbf{J}$ contains drift and diffusion components:
$\mathbf{J}= \mathbf{J}_\mathrm{drift} + \mathbf{J}_\mathrm{diff}$.

The IX diffusion current $\mathbf{J}_\mathrm{diff}$ characterized by the diffusion coefficient $D$ is given by:
\begin{equation}
\mathbf{J}_\mathrm{diff}=
-D \nabla n.
\label {eq:diffcurr}
\end{equation}
Here we use the value $D=0.4$~cm$^2$/s consistent with our previous work on the similar samples studied under similar conditions \cite{Chiaruttini2019,Chiaruttini2021}. 
 
The drift term $\mathbf{J}_\mathrm{drift}$ in Eq.~\ref{eq:drift-diffu} can be
written as
\begin{equation}
\mathbf{J}_\mathrm{drift}=
-\mu n \nabla (\phi_0 n+ U(x,y)),
\label{eq:Jdrift}
\end{equation}
where  $\mu={D}/({k_{B}T})$ is the IX mobility and $k_B$ is the Boltzmann constant. This current is due to both exciton-exciton interaction energy $\phi_0 n$ and an external potential $U(x,y)=U_{t0} U_{xy}$ imprinted by the electrodes. Here $U_{t_0}$ is the trap depth, and $U_{xy}$ is normalized to unity. 
%

The recombination rate $R$ can be expressed in terms of  radiative  $\tau_{rad}$ and non-radiative $\tau_{nrad}$ recombination times:
\begin{equation}
R= \frac{1}{\tau_{r}}+ \frac{1}{\tau_{nr}}.
\end{equation}
The density dependent radiative recombination time is given by $\tau_{r}=\tau_{0r} \exp ( - n/\gamma)$, where 
$\tau_{r0}=1$~$\mu$s is estimated from our previous work on similar samples \cite{Fedichkin2016}  and  $\gamma$ is determined as described in Appendix~\ref{sec:appendix_param}. 
The non-radiative recombination rate is assumed to depend exponentially on the absolute value of the  in-plane component of the electric field  $F_{xy} = \nabla U(x,y)$. 
\begin{equation}
\frac{1}{\tau_{nr}}=   \frac{1}{\tau_{nr0}} \mathrm{exp}(\frac{\lVert F_{xy}\rVert }{F_0}),
\end{equation}
%
{where $\tau_{nr0}$ is the non-radiative recombination time in the absence of the in-plane electric field, $F_0$  is a critical field leading to IX dissociation.  The latter can be roughly estimated as a ratio between IX binding energy and in-plane Bohr radius, but {\it a priori} their values are unknown, and may be used as fitting parameters. }
%

%

%


%

To solve numerically the drift-diffusion equation~\ref{eq:Jdrift},
we proceed into two steps.

First, the in-plane potential $U(x,y)$  and the corresponding in-plane electric field $F_{xy}$ are
calculated numerically by solving in three dimensions the Poisson equation.
%
It turns out that close to the trap edge $U_{xy}$ is well approximated by the error function:
\begin{equation}
U_{xy}= \frac{1}{2}
\left[ 
\mathrm{erf}
\left( -\frac{r(x,y)}{\sqrt{2} \sigma}
\right)+1\right], 
\label{eq:edge}
\end{equation}
where
$r(x,y)$ is the distance between the in-plane position ($x,y$) and the trap edge (using the relative normal coordinate) 
and $\sigma \simeq $ 100 nm is the standard deviation of the error function.
The value of $\sigma$ depends essentially on the distance between the QW and the top electrode.
Note also  that, for simplicity, this model neglects the density dependence of the trapping potential.

The second step consists in using a two-dimensional finite element method \cite{freefem} 
to solve numerically Eq.~\ref{eq:Jdrift}, 
in which Eq.~\ref{eq:edge} is injected  considering $\sigma$ as an additional parameter. The best agreement with experimental results is obtained by increasing $\sigma$ up to $2$~$\mu$m. %
This reveals that 
the model overestimates the non-radiative recombination rate at high electric fields.
We speculate, that accounting for the density dependence of $\tau_{nr}$ could amend the agreement with the in-plane potential profile calculated from the Poisson equation. 
The values of over fitting parameters are  $N_p=4\times10^{11}$~s$^{-1}$, $\tau_{nr0}=1$~$\mu$s and $F_0=30$~kV/cm.
%
%
%
%

The calculated steady-state solutions $n(x,y)$  of  the non-linear equation (\ref{eq:drift-diffu})  
are compared to the experimentally determined density profiles. 
Fig.~\ref{fig:freefem}~(a) shows IX density profiles $n(x,y=0)$  corresponding to the data shown in Fig.~\ref{fig:ColormapInsideHigh}~(b)-(d), while  Fig.~\ref{fig:freefem}~(b) shows the results of the modelling.  
The comparison between measured and calculated integrated IX densities in the trap  at different values of the applied electric bias is shown in Fig.~\ref{fig:freefemAllVb}.
%
%

The qualitative agreement between the model and the data can be ascertained. Namely, (i) strong reduction of IX population in the trap at $V_b=+0.8$~V as compared to the absence of an external bias; (ii) a weak reduction of this population at $V_b=-0.8$~V.
The reduction of the trap depth from $U_{t0}=60$~meV at $V_b=-0.8$~V to $U_{t0}=0$~meV at $V_b=+0.8$~V is consistent with the experimental result $R_e=38$~meV/V, see Fig.~\ref{fig:VbE}.

However, the quantitative agreement could not be reached. In particular, the model yields much stronger reduction of the IX density at negative bias than observed experimentally. Elaboration of a more complex model taking account of density-dependence of the in-plane electric field could be relevant for a better understanding of IX trapping and release.

\bibliography{biblio}

\end{document}